\documentclass[aps,prb,twocolumn,superscriptaddress,print]{revtex4-2}
\usepackage{amssymb,amsmath}
\usepackage{graphicx}
\usepackage{multirow}
\usepackage{hyperref}
\usepackage{appendix}
\usepackage{mathtools}
\usepackage{amsmath}
\usepackage{verbatim}
\usepackage{gensymb}
\usepackage{subfigure}
\usepackage{bm}
\usepackage{threeparttable}
\usepackage{booktabs}
\usepackage{braket}
\usepackage{xcolor}
\usepackage[normalem]{ulem}
\usepackage[mathscr]{euscript}
\usepackage{tabularx}

\begin{document}
	
\title{Plasmon excitations in monolayer stanene}
	
\author{Jia-Hui Qin}
\affiliation{Lanzhou Center for Theoretical Physics, Key Laboratory of Quantum Theory and Applications of the Ministry of Education, Key Laboratory of Theoretical Physics of Gansu Province, Gansu Provincial Research Center for Basic Disciplines of Quantum Physics, School of Physics and Technology, Lanzhou University, Lanzhou 730000, China}

\author{Yunhua Wang}
\email{wangyunhua@lzu.edu.cn}
\affiliation{Lanzhou Center for Theoretical Physics, Key Laboratory of Quantum Theory and Applications of the Ministry of Education, Key Laboratory of Theoretical Physics of Gansu Province, Gansu Provincial Research Center for Basic Disciplines of Quantum Physics, School of Physics and Technology, Lanzhou University, Lanzhou 730000, China}
	
	
\begin{abstract}
Monolayer stanene exhibits strong intrinsic spin-orbit coupling (SOC) and notably distinct low-energy band features involving both $\Gamma$ and K valleys together with strong electron-hole asymmetry among group-IV monolayers. Its plasmonic properties and external-field tunability remain largely unclarified. Here, we systematically investigate its high-energy plasmon structures and low-energy collective excitation modulated by temperature, electrostatic doping and perpendicular electric field. Calculations use a $sp^3$ tight-binding model with SOC and sublattice staggered potentials, combined with the finite-temperature multiband Lindhard function under the random-phase approximation. Besides the $\pi$ and $\pi+\sigma$ plasmon known in graphene/silicene, stanene supports two new high-energy plasmon branches. For low-energy plasmons, thermal excitation extends the accessible energy-momentum range while preserving high quality factors and collective coherence. Electrostatic doping yields distinct plasmon behaviors: moderate doping gives K-valley dominated intraband $\sqrt{q}$ Dirac plasmons; heavy electron doping retains $\sqrt{q}$ dispersion via Drude contributions from $\Gamma$ and K valleys; heavy hole doping triggers interband transitions near the $\Gamma$ point, producing new $\sigma$-$\sigma$ and $\sigma$-$\sigma^*$ interband plasmon modes and complex multi-branch spectra. Furthermore, the perpendicular electric field enables continuous tuning of the $\sqrt{q}$ plasmon via band reconstruction below 0.3 V/Å and induces extra plasmon modes near band-gap closure. These tunable low-energy plasmons span terahertz-to-mid-infrared frequencies, promising for nanophotonic applications.
\end{abstract}

\maketitle

\section{Introduction}

As a member of the two-dimensional (2D) group-IV material family, stanene possesses distinctive properties different from graphene, featured by a characteristic buckled atomic geometry, strong spin-orbit coupling (SOC), and highly sensitive surface-mediated interactions. Benefiting from these remarkable physical features, stanene can exhibit various exotic topological quantum states \cite{zhao2020stanene,molle2017buckled,liu2011low,xu2013large,ohtsubo2013dirac,zhang2014honeycomb,wu2014prediction,tang2014stable,
ezawa2015mono,deng2018epi,liao2018sup,falson2020type,zhao2022coe}, prominent thermoelectric performance \cite{xu2014enhanced,zhou2016quantum}, as well as highly tunable electronic and magnetic properties through various external modulation strategies: chemical functionalization \cite{xu2013large,wang2014two,chou2014hydro,xu2024second,rani2023stanene}, substrate heterostructure engineering \cite{zhu2015epitaxial,sahoo2019per,zang2022molecular,ochapski2022pro,zhang2016quantum,ni2017ger,xu2018gapped,di2019towards,
zheng2020epi,yuhara2021plane,barman2024growth,eltinge2022epi,li2024mult,wang2025quantum}, strain engineering \cite{wang2015effect,gou2017strain,lu2017quasi,liu2019real}, out-of-plane electric-field modulation \cite{ezawa2015mono,houssa2016top}, or magnetic-field modulation \cite{rachel2014giant,yu2022response}. Compared with silicene and germanene, stanene exhibits a larger buckling height and stronger SOC. Apart from the low-energy gapped Dirac bands originating from the $p_z$ orbitals near the K and K$^\prime$ points, where the SOC opens a band gap of approximately 0.11 eV, the energy bands at the $\Gamma$ point lie close to the Fermi level for stanene \cite{xu2013large}. Upon decoration of stanene with chemical functional groups or substrates, the $p_z$ orbitals become saturated and are shifted away from the Fermi level by several eVs. In this case, the bands at the $\Gamma$ point dominate the electronic behavior. In recent years, both monolayer and few-layer stanene on various substrates have been successfully fabricated, and their topological quantum states experimentally measured, enabled by state-of-the-art molecular-beam epitaxy growth techniques \cite{zhu2015epitaxial,sahoo2019per,deng2018epi,liao2018sup,falson2020type,zhao2022coe,zang2022molecular,ochapski2022pro,
xu2018gapped,di2019towards,zheng2020epi,yuhara2021plane,barman2024growth}.

Plasmons, as an essential many-body effect, refer to the quantized collective density oscillations of an electron liquid. Plasmons in 2D Group-IV materials with special emphasis on graphene and silicene have fundamental and practical importance for surface plasmon physics and technology \cite{wang2011found,das2011electronic,basov2011elec,kotov2012electron,tame2013quantum,
fei2012gate,grigorenko2012g,bao2012graphene,luo2013plasmons,
jablan2013plasmons,low2014graphene,basov2014colloquium,
xia2014two,stauber2014plas,politano2014plasmon,garcia2014graphene,
xiao2016graphene,thygesen2017cal,huang2017graphene,fan2019graphene,
rodin2020collective,li2020graphene,in2020dirac,elbanna20232d,kuang2024optical}. Compared with the conventional 2D electron gas, graphene hosts 2D Dirac plasmons \cite{wunsch2006dynamical,hwang2007dielectric,wang2007collective,polini2008plasmons,das2009collective}, exhibiting the typical characteristic $\sqrt{q}$ dispersion, yet with an unusual carrier-density ($n$) scaling of $n^{1/4}$, different from the $\sqrt{n}$ dependence in the conventional 2D electron gas. Compared with conventional metallic surface plasmonics, graphene plasmonics work in lower frequency ranges varying from terahertz to mid-infrared frequencies and enable novel applications in terahertz optoelectronics \cite{fei2012gate,grigorenko2012g,bao2012graphene,luo2013plasmons,
jablan2013plasmons,low2014graphene,basov2014colloquium,
xia2014two,stauber2014plas,politano2014plasmon,garcia2014graphene,
xiao2016graphene,thygesen2017cal,huang2017graphene,fan2019graphene,
rodin2020collective,li2020graphene,in2020dirac}. Furthermore, graphene plasmonics have the advantages of strong sub-wavelength confinement, low power consumption, long lifetimes, extremely high speed, and high external-field tunability including electric field \cite{mishchenko2010guided,zhang2022electric}, magnetic field \cite{poumirol2013mag}, SOC \cite{wang2007collective,scholz2012d}, gate-controllable doping \cite{wunsch2006dynamical,hwang2007dielectric,wang2007collective,polini2008plasmons,das2009collective}, strain \cite{pellegrino2010d,principi2016edge}, and substrate surface phonon \cite{liu2010plasmon,hwang2010plasmon,koch2010strong,jablan2011un}. In addition to the Dirac plasmons, graphene has higher energy $\pi$ and $\pi+\sigma$ plasmons arising from the interband processes between $\pi$ and $\sigma$ bands \cite{politano2014plasmon,wachsmuth2013high,liou2015pi}. Silicene and germanene have corresponding SOC-induced band gaps with about 4 and 40 meV as well as a sublattice-dependent staggered potential induced by a perpendicular electric field. Their low-energy electronic behaviors are mainly governed by the effective Hamiltonian near $\mathrm{K}$ and $\mathrm{K}^\prime$ valleys \cite{molle2017buckled}, and topological phase transitions among quantum spin Hall states, valley-polarized anomalous Hall states, spin- and valley-polarized metals, and band insulators can be generated by an external electric field or an exchange field \cite{liu2011quantum,pan2014valley,ezawa2012valley,bampoulis2023q}. The undamped 2D Dirac plasmon modes exhibit a remarkable external-field tunability and some special features for different quantum phases outside the particle-hole continuum regions \cite{tabert2014dynamical,chang2014probing,van2014spin,xue2023electronic,rudenko2024ani,vlzvjpq7,elgvin2026a}. Furthermore, temperature, electric field, magnetic field, and polar substrate have a remarkable influence on the susceptibility and Dirac plasmon mode of silicene and germanene \cite{vacacela2017plasmon,li2017first,iurov2017thermal,shih2018coulomb,Mirzaei2018,sarkar2019dynamical}. The higher energy $\pi$ and $\pi+\sigma$ plasmons also present a distinctive double-peak structure in silicene and germanene unlike graphene \cite{vacacela2017plasmon,li2023first}.

\begin{figure*}[t]
	\centering
	\includegraphics[width=14cm]{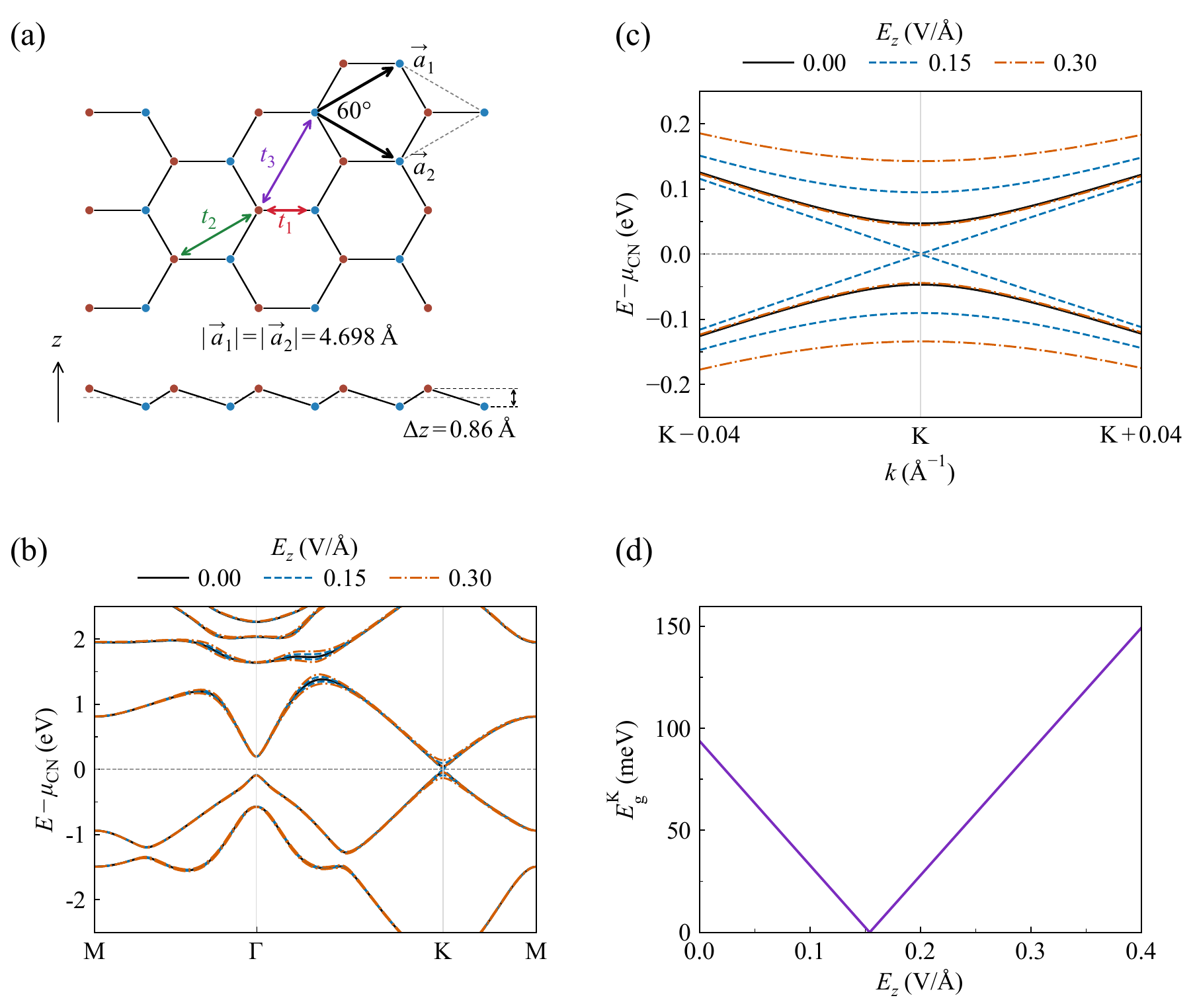}
	\caption{
    (a) Top and side views of the buckled honeycomb lattice of stanene, with the primitive lattice vectors $\vec{a}_1$ and $\vec{a}_2$, the first-, second-, and third-nearest-neighbor hopping terms $t_1$, $t_2$, and $t_3$, respectively, and the buckling height $\Delta z$. (b) Band structures along the $\mathrm{M}$--$\Gamma$--$\mathrm{K}$--$\mathrm{M}$ path for
    $E_z=0$, $0.15$, and $0.30$ V/\AA, respectively. (c) A zoom-in view of the low-energy bands near the $\mathrm{K}$ valley in (b). (d) The band gap $E^K_g$ as a function of the perpendicular electric-field $E_z$.
	}
	\label{fig1}
\end{figure*}

Compared with graphene, silicene and germanene, whose low-energy physics is dominated by Dirac-like Hamiltonian at $\mathrm{K}$ and $\mathrm{K}^\prime$ valleys, stanene has the strongest SOC with a band gap about 110 meV as well as apparently distinctive low-energy band features, i.e., bands near both $\Gamma$ and $K$ valleys lie close to the Fermi level \cite{xu2013large,pedersen2010tight,hattori2017edge,jiang_tight_2017,Salazar2017,Fuhr2021}. It is naturally anticipated that stanene has distinctive plasmon structures and external-field tunability. Despite extensive efforts devoted to collective plasmon excitations in graphene and silicene, the high-energy plasmon structures and the effects of finite temperature, electrostatic doping, and perpendicular electric fields on many-body dynamical responses and low-energy plasmon spectra of stanene remain largely unexplored. Herein, we adopt the tight-binding (TB) model with $\{s, p_x, p_y, p_z\}$ as bases incorporating spin–orbit coupling and field-induced sublattice stagger potentials, combined with the finite-temperature multiband Lindhard function and random-phase approximation, to systematically characterize the high-energy plasmon features as well as low-energy collective plasmon spectra of monolayer stanene. We find that stanene has four distinguished high-energy plasmon structures. We examine how thermal effects, chemical potential shifts, and vertical electric fields tune dynamical electronic responses and low-energy plasmon spectra. Thermal excitation expands the accessible phase space for plasmon modes and preserves robust collective coherence over a broad temperature window. Electrostatic gating shifts the chemical potential and redistributes Fermi-sea carriers, activating several interband transitions: subband–subband transitions within the valence manifold and valence-to-conduction band transitions near the $\Gamma$ point. Depending on doping types and strengths, distinct plasmon behaviors emerge: moderate doping favors the $\sqrt{q}$ Dirac plasmon originating from the intraband transition of $\mathrm{K}$-point Dirac bands, heavy electron doping sustains the $\sqrt{q}$ dispersion via parallel Drude-type intraband contributions from both $\Gamma$ and $\mathrm{K}$ valleys, whereas heavy hole doping generates new $\sigma$-$\sigma$ and $\sigma$-$\sigma^*$ interband plasmon modes and complex multi-branch plasmon dispersions. Vertical electric field $E_z$ reconstructs the band structure and redistributes thermally populated carriers, yielding continuous tunability of the $\sqrt{q}$ plasmon within 0 V/{\AA} $< E_z < 0.3$ V/{\AA} and producing additional plasmon branches near the band-gap vanishing point. These highly tunable low-energy plasmons, residing in the terahertz to mid-infrared regime, together with good compatibility with conventional field-effect-transistor fabrication protocols, render stanene an attractive candidate for future terahertz and mid-infrared nanophotonic applications.

\begin{table}[t]
	\centering
    \caption{Parameters of the $sp^{3}$ tight-binding model used in the present calculations \cite{jiang_tight_2017}, where all energy parameters are in eV.
	}
	\label{tab:stanene_tb}
	\setlength{\tabcolsep}{4.8mm}{
	\begin{tabular}{l r @{\hspace{1.5em}} l r}
		\hline\hline
		Parameter & Value & Parameter & Value \\
		\hline
		
		$\varepsilon_s$
		& $-5.1576$
		& $\varepsilon_p$
		& $0.4728$
		\\
		
		$V_{ss\sigma}^{(1)}$
		& $-1.2531$
		& $V_{sp\sigma}^{(1)}$
		& $1.8809$
		\\
		
		$V_{pp\sigma}^{(1)}$
		& $1.5222$
		& $V_{pp\pi}^{(1)}$
		& $-0.7384$
		\\
		
		$V_{ss\sigma}^{(2)}$
		& $-0.0496$
		& $V_{sp\sigma}^{(2)}$
		& $-0.0358$
		\\
		
		$V_{pp\sigma}^{(2)}$
		& $0.1020$
		& $V_{pp\pi}^{(2)}$
		& $-0.0236$
		\\
		
		$V_{ss\sigma}^{(3)}$
		& $0.0537$
		& $V_{sp\sigma}^{(3)}$
		& $0.0507$
		\\
		
		$V_{pp\sigma}^{(3)}$
		& $0.1341$
		& $V_{pp\pi}^{(3)}$
		& $-0.0010$
		\\
		
		$\Delta_{\mathrm{SO}}$
		& $0.672$
		&
		&
		\\
		
		\hline\hline
	\end{tabular} }
\end{table}

\section{Model and methods}
\label{Methods}

Monolayer stanene consists of a buckled honeycomb lattice with two Sn sublattices in a unit cell in Fig.~\ref{fig1}(a). In contrast to the planar honeycomb lattice of graphene, the two sublattices are displaced oppositely along the $z$ direction. We consider the relaxed structural geometry of stanene, with the lattice constant $a_{0}=4.698~\text{\AA}$ and the buckling height $\Delta z=0.86~\text{\AA}$, and the electronic behavior of pristine stanene can be well captured by the TB model with the $\{s,p_x,p_y,p_z\}$ orbital basis of each Sn atom~\cite{jiang_tight_2017}. In this work, we introduce a perpendicular electric-field $E_z$ to consider the field-induced tunability of its electronic structure and plasmon. The total TB Hamiltonian is written as
\begin{equation}
	\begin{aligned}
		H ={}&
		\sum_{i\alpha\sigma}
		\varepsilon_{\alpha}
		c_{i\alpha\sigma}^{\dagger}
		c_{i\alpha\sigma}
		\\
		&+
		\sum_{n=1}^{3}
		\sum_{\langle i,j\rangle_n}
		\sum_{\alpha\beta\sigma}
		\left[
		t_{\alpha\beta}^{(n)}
		(\hat{\mathbf d}_{ij})
		c_{i\alpha\sigma}^{\dagger}
		c_{j\beta\sigma}
		+\mathrm{H.c.}
		\right]
		\\
		&+
		\frac{\Delta_{\mathrm{SO}}}{3}
		\sum_i
		\mathbf{L}_i\cdot\boldsymbol{\sigma}_i
		+
		\sum_{i\alpha\sigma}
		U_{\xi_i}
		c_{i\alpha\sigma}^{\dagger}
		c_{i\alpha\sigma}.
	\end{aligned}
	\label{eq:stanene_hamiltonian}
\end{equation}
Here, $i$ and $j$ denote Sn sites, $\alpha,\beta\in\{s,p_x,p_y,p_z\}$ label the atomic orbitals, and $\sigma$ denotes spin. The notation $\langle i,j\rangle_n$ represents the $n$th-nearest-neighbor site pairs, with $n=1$, $2$, and $3$ corresponding to the first, second, and third nearest neighbors, respectively. In the first term $\varepsilon_\alpha$ denotes the on-site energy, and in the second term the hopping amplitudes $t_{\alpha\beta}^{(n)}(\hat{\mathbf d}_{ij})$ are constructed from the Slater-Koster two-center integrals $V_{ss\sigma}^{(n)}$, $V_{sp\sigma}^{(n)}$, $V_{pp\sigma}^{(n)}$, and $V_{pp\pi}^{(n)}$ \cite{Slater-Koster}, with the bond vector $\hat{\mathbf d}_{ij}$.

The third term in Eq.~\eqref{eq:stanene_hamiltonian} describes the on-site atomic spin-orbit coupling (SOC) within the $p$-orbital subspace, i.e., $\{|p_x,\uparrow\rangle, |p_y,\uparrow\rangle, |p_z,\uparrow\rangle, |p_x,\downarrow\rangle, |p_y,\downarrow\rangle, |p_z,\downarrow\rangle\}$, where $\mathbf{L}$ is the orbital angular-momentum operator and $\boldsymbol{\sigma}$ denotes the Pauli matrices acting on spin, and $\Delta_{\mathrm{SO}}$ is the SOC splitting parameter. The $s$ orbital does not contribute to the on-site SOC because its orbital angular momentum vanishes. All the on-site, hopping, and SOC parameters used in this work are summarized in Table~\ref{tab:stanene_tb}~\cite{jiang_tight_2017}.

The last term in Eq.~(1) represents the contribution from the perpendicular electric-field $E_z$. Since the two Sn sublattices are displaced along the out-of-plane direction, the field produces opposite on-site energy shifts on the two sublattices. We take the upper and lower sublattices to be located at $z_A=+\Delta z/2$, and $z_B=-\Delta z/2$, respectively, and define positive $E_z$ along the $+z$ direction. The corresponding on-site energy shifts with $\xi_i = A$ or $B$ are therefore $U_A=+eE_z\Delta z/2$, and $U_B=-eE_z\Delta z/2$, where $e>0$ is the elementary charge. Here $E_z$ is expressed in V/\AA{} and the energy shift is expressed in eV. The reconstructed band structures of stanene by $E_z$ are shown in Figs.~\ref{fig1}(b)-\ref{fig1}(d) with band gap closing and reopening accompanying a topological quantum phase transition \cite{ezawa2012valley,van2014spin,tabert2014dynamical}.

The noninteracting density response was evaluated by using the finite-temperature multiband Lindhard function \cite{giuliani2005quantum},
\begin{equation}
	\begin{aligned}
		\chi(\mathbf q,\omega)
		={}&
		\frac{g_s}{(2\pi)^2}
		\int_{\mathrm{BZ}} d^2\mathbf k
		\sum_{n,m}
		\\
		&\times
		\frac{
			f(E_{n\mathbf k})-
			f(E_{m,\mathbf k+\mathbf q})
		}{
			E_{n\mathbf k}
			-
			E_{m,\mathbf k+\mathbf q}
			+
			\hbar\omega+i\eta
		}
		\\
		&\times
		\left|
		\left\langle
		m,\mathbf k+\mathbf q
		\left|
		e^{i\mathbf q\cdot\mathbf r}
		\right|
		n,\mathbf k
		\right\rangle
		\right|^2 .
	\end{aligned}
	\label{eq:lindhard}
\end{equation}

Here $E_{n\mathbf k}$ and $|n,\mathbf k\rangle$ are the tight-binding eigenvalues and eigenstates, respectively, $\eta>0$ is the finite energy broadening entering the retarded response, and $f(E)=\{\exp[(E-\mu)/(k_{\mathrm B}T)]+1\}^{-1}$ is the Fermi-Dirac distribution. The broadening of $\eta=5~\mathrm{meV}$ is used throughout the dynamical-response calculations. Equation~(\ref{eq:lindhard}) is written in its conventional form with $\hbar$ shown explicitly. Throughout the numerical calculations, we set $\hbar=1$, so that $\omega$ is expressed in energy units and $\eta$ has the same energy dimension as $\omega$.
Since spin is already included explicitly in the 16-band Hamiltonian, $g_s=1$ is used in the response calculation. With the two-dimensional normalization adopted in Eq.~(\ref{eq:lindhard}), $\chi(q,\omega)$ is an areal polarization function with units of $\mathrm{eV}^{-1}\mathrm{nm}^{-2}$.

\begin{figure*}[t]
	\centering
	\includegraphics[width=16cm]{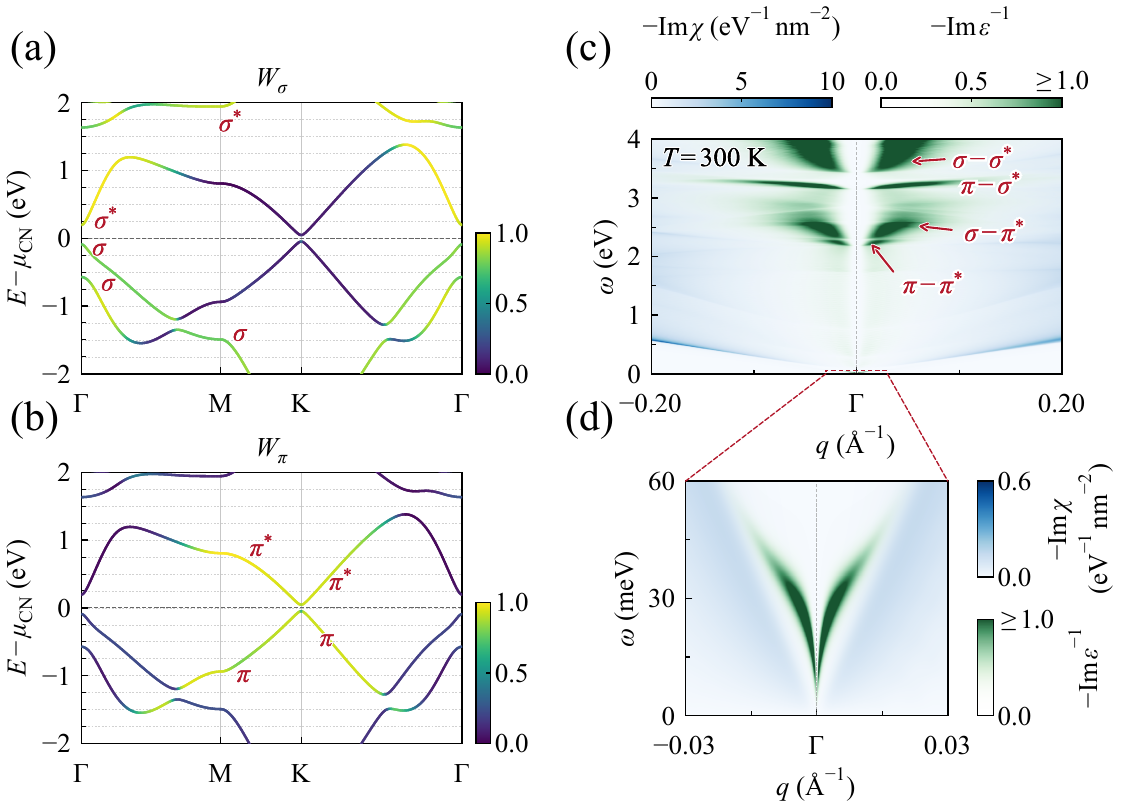}
    \caption{Projected electronic band structures into (a) $\sigma$ and (b) $\pi$ electron hopping channels within the TB Slater-Koster model, where $W_\sigma$ and $W_\pi$ are their corresponding channel weights with $W_\sigma+W_\pi = 1$. In the eigenband space, energy bands are labeled as $\pi$, $\pi^*$, $\sigma$, or $\sigma^*$ bands via $W_\sigma$ and $W_\pi$,    
    with the superscript asterisk denoting conduction bands. (c) Momentum- and energy-resolved noninteracting response function $-\operatorname{Im}\chi(q,\omega)$ and energy-loss function $-\operatorname{Im}[1/\varepsilon(q,\omega)]$ as functions of momentum transfer $\mathbf{q}$ and excitation energy $\omega$
    along M--$\Gamma$ (left) and $\Gamma$--K (right) at $T=300$ K, where high-energy $\pi$-$\pi^*$, $\sigma$-$\pi^*$, $\pi$-$\sigma^*$, and $\sigma$-$\sigma^*$ plasmon modes are denoted. (d) A zoom-in view of the low-energy Dirac plasmon near $\Gamma$ point.
	}
	\label{fig:full_spectrum}
\end{figure*}

In the numerical calculations, the chemical potential is defined as the charge-neutrality energy point, for each temperature $T$ and
each perpendicular electric field $E_z$, namely, $\mu_{\mathrm{CN}}(T,E_z)$, determined by imposing the charge-neutrality condition, 
$n_e(\mu_{\mathrm{CN}},T,E_z) = n_h(\mu_{\mathrm{CN}},T,E_z)$, where $n_e$ and $n_h$ denote the electron and hole carrier densities,
respectively, obtained from the Fermi-Dirac occupations of the conduction and valence bands. If a back gate is applied, the chemical potential $\mu$ can be tuned by the back gate voltage. Then the chemical potential $\mu$ is written as $\mu=\mu_{\mathrm{CN}}(T,E_z)+\Delta\mu$. Thus, $\Delta\mu=0$ corresponds to the charge neutrality of undoped stanene, while $\Delta\mu>0$ and $\Delta\mu<0$ correspond to electron and hole dopings, respectively.

Electron-electron screening is treated within the random-phase approximation, and in this respect the dielectric function is expressed by
\begin{equation}
	\varepsilon(\mathbf q,\omega)
	=
	1-
	V(q)\chi(\mathbf q,\omega).
	\label{eq:dielectric}
\end{equation}
For a two-dimensional system, the Fourier-transformed Coulomb interaction is taken as $V(q)=e^2/(2\varepsilon_0\varepsilon_r q)$, where $\varepsilon_r$ is the relative dielectric constant of background. For a freestanding system in vacuum, we set $\varepsilon_r=1$. The collective charge response is characterized by the following energy-loss function \cite{Slotman2018,Kuang2021,Feng2025}
\begin{equation}
	\begin{aligned}
		\mathcal{L}(\mathbf q,\omega)
		&=
		-\operatorname{Im}
		\left[
		\frac{1}{\varepsilon(\mathbf q,\omega)}
		\right]
		\\
		&=
		\frac{
			\operatorname{Im}\varepsilon(\mathbf q,\omega)
		}{
			\left[\operatorname{Re}\varepsilon(\mathbf q,\omega)\right]^2
			+
			\left[\operatorname{Im}\varepsilon(\mathbf q,\omega)\right]^2
		}.
	\end{aligned}
	\label{eq:loss_function}
\end{equation}

Here, $\operatorname{Re}\varepsilon(\mathbf q,\omega)$ and $\operatorname{Im}\varepsilon(\mathbf q,\omega)$ denote the real and imaginary parts of the dielectric function, respectively. A well-defined plasmon resonance is typically associated with a zero crossing of $\operatorname{Re}\varepsilon(\mathbf q,\omega)$ accompanied by a sufficiently small $\operatorname{Im}\varepsilon(\mathbf q,\omega)$ in Eq.~(\ref{eq:loss_function}). In this respect, the energy-loss function exhibits a sharp resonance peak, which approaches a delta-function-like form in the ideal lossless limit. The peak position characterizes the plasmon energy, whereas its spectral width reflects the damping of the collective excitation.

To quantify the spectral sharpness of a plasmon resonance, we use the full width at half maximum (FWHM) of the loss-function peak, which includes the intrinsic and extrinsic parts induced by the Landau damping $\gamma$ and the imposed broadening $\eta$, respectively. The corresponding quality factor $Q_{\mathbf q}$ is here defined as,
\begin{equation}
	Q_{\mathbf q}=\frac{E_p}{\mathrm{FWHM}-2\eta},
\label{eq:qf}
\end{equation}
where $E_p=\hbar\omega_p(\mathbf q)$ denotes the plasmon peak energy. The dimensionless quality factor $Q_{\mathbf q}$ in Eq.~(\ref{eq:qf}) measures the plasmon peak energy in units of the intrinsic Landau damping, namely, $2\gamma=\mathrm{FWHM}-2\eta$~\cite{bosman2013}. A larger $Q_{\mathbf q}$ means that the intrinsic linewidth is relatively small compared with the resonance energy, and hence the plasmon is better defined spectrally. The constructions and calculations of tight-binding Hamiltonian, Lindhard function, and dielectric function are realized in the numerical TBPLaS package~\cite{LI2023108632}.

\section{Results and Discussion}
\label{Results and Discussion}

\subsection{High-energy plasmons}
Figures~\ref{fig:full_spectrum}(a) and \ref{fig:full_spectrum}(b) show the projected band structures of monolayer stanene with respect to $\pi$, $\pi^*$, $\sigma$, and $\sigma^*$ bands, with the superscript asterisk denoting conduction bands. Here, within the TB Slater-Koster model, electron hopping between two atoms along a chemical bond splits into two categories, i.e., $\sigma$ and $\pi$ hopping channels. In the eigenband space, conduction and valence bands can be mapped on the two hopping channels. Therefore, according to their corresponding channel weights $W_\sigma$ and $W_\pi$, we can label these energy bands as $\pi$, $\pi^*$, $\sigma$, or $\sigma^*$ bands. Near the $\mathrm{K}$ point, the Dirac bands consist of $\pi$ and $\pi^*$ bands from the $p_z$ orbital similar to that of graphene, silicene, and germanene. The collective coherent superposition of intraband electron-hole pair excitations near the Fermi surface of the doped Dirac bands under Coulomb interaction enables the low-energy Dirac plasmons. Near the $\Gamma$ point, the bands contributed by the $\sigma$ and $\sigma^*$ bands are also near the Fermi level, which is absent in graphene, silicene, and germanene. Consequently, besides the Dirac plasmon, new low-energy plasmon modes with $\sigma \rightarrow \sigma^*$ electron-hole excitation channel might emerge in stanene. At the $\mathrm{M}$ point, besides the $\pi$ and $\pi^*$ bands allowing the same $\pi \rightarrow \pi^*$ transition as that of silicene, the $\sigma$ and $\sigma^*$ bands with nearby energies in stanene can enable additional $\pi \rightarrow \sigma^*$, $\sigma \rightarrow \pi^*$, and $\sigma \rightarrow \sigma^*$ transitions. As a result, it is anticipated that several new high-energy plasmon modes emerge possibly with nearby energies of $\pi$ plasmon in stanene.

Figure~\ref{fig:full_spectrum}(c) presents the noninteracting response function $-\operatorname{Im}\chi(\mathbf q,\omega)$ expressed in units of $\mathrm{eV}^{-1}\mathrm{nm}^{-2}$ as well as the energy-loss function $\mathcal{L}(\mathbf q,\omega) = -\operatorname{Im}[1/\varepsilon(\mathbf q,\omega)]$, as functions of momentum transfer $\mathbf q$ and excitation energy $\omega$, along the M--$\Gamma$--K path, within a broader energy range of $\omega$ spanning $0$-$4~\mathrm{eV}$, at $T=300$ K. From $-\operatorname{Im}\chi(\mathbf q,\omega)$ (blue color bar), we can see remarkable Lindhard continuum $[-\operatorname{Im}\chi(\mathbf q,\omega)>0]$, i.e., single-particle electron-hole excitation region with momentum and energy conservations. The energy-loss function $\mathcal{L}(\mathbf q,\omega)$ (green color bar) reveals collective plasmon excitations, which appear as bright and high-intensity regions denoted by the deep green. This map clearly illustrates the fundamental competition between collective plasmon excitations and single-particle electron–hole excitations in the 2D stanene within a broader energy range. The well-defined plasmon branch survives only outside the Lindhard continuum, where the very weak Landau damping is negligible. Upon entering the single-particle continuum, the plasmon suffers strong Landau damping, leading to peak broadening and attenuation of the resonance intensity.

Within the energy range about from 2 to 4 eV in Fig.~\ref{fig:full_spectrum}(c), four plasmon structures can be clearly distinguished: (i) the $\pi$-$\pi^*$ plasmon, i.e., the so-called $\pi$ plasmon, with energy range from about 2.15 to 2.3 eV, (ii) the $\sigma$-$\pi^*$ plasmon mode, with energy range from about 2.35 to 2.65 eV, (iii) the $\pi$-$\sigma^*$ plasmon mode with energy range from about 3.15 to 3.35 eV, and (iv) the $\sigma$-$\sigma^*$ plasmon mode with energy larger than about 3.4 eV. Comparing with the $\pi$ and $\pi+\sigma$ plasmon modes in graphene, silicene, and germanene \cite{politano2014plasmon,wachsmuth2013high,liou2015pi,vacacela2017plasmon,li2023first}, two new plasmon branches with distinguished energy ranges emerge in stanene. 

\begin{figure*}[t]
	\centering
	\includegraphics[width=\textwidth]{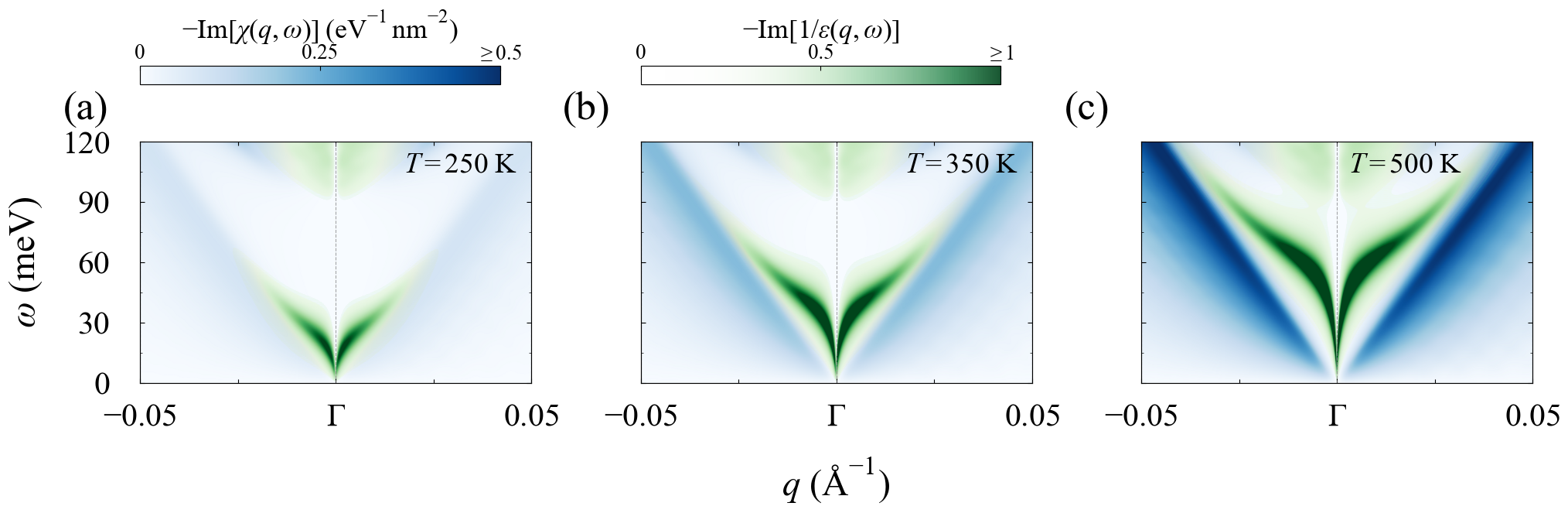}
	\caption{
       Low-energy dynamical response and energy-loss function of charge-neutral stanene at $T=250$, $350$, and $500~\mathrm{K}$ in (a)-(c), respectively. 
	}
	\label{fig:temperature_spectra}
\end{figure*}

 \begin{figure*}[t]
 	\centering
 	\includegraphics[width=\textwidth]{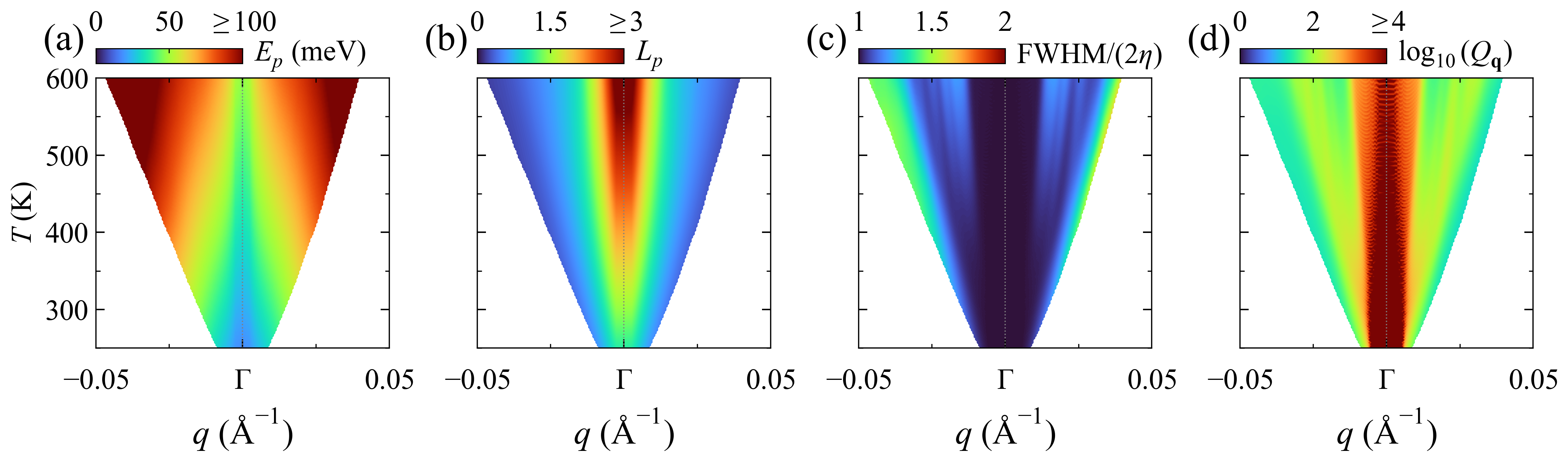}
 	\caption{
 		$(\boldsymbol q, T)$ maps of the low energy plasmon peak energy $E_p$, peak height $L_p$, normalized linewidth
 		$\mathrm{FWHM}/(2\eta)$, and quality factor $Q_{\boldsymbol q}$ in (a)-(d) for charge-neutral stanene, respectively.
 	}
 	\label{fig:temperature_phase}
 \end{figure*}

The in-plane intraband plasmon predominantly falls within the low-energy terahertz to far-infrared range near the $\Gamma$ point in Fig.~\ref{fig:full_spectrum}(d). Since the collective mode resides outside the Lindhard single-particle continuum, the Landau damping channel is largely suppressed. The small value of $-\operatorname{Im}\chi(\mathbf q,\omega)$ indicates few available electron–hole pair decay pathways, and hence the plasmon maintains a sharp resonance. Studying this low-energy plasmon of 2D stanene near $\Gamma$-point is of particular importance for several reasons. First, the long-wavelength limit ($q\rightarrow 0$) with $q = |\mathbf q|$ is directly relevant to experimental far-field optical excitation, because incident photons carry very small in-plane momentum. Second, 2D stanene like other 2D materials also exhibits the characteristic $\omega_p \propto \sqrt{q}$, which distinguishes them from 3D bulk plasmons with nearly constant plasma frequency. Third, the terahertz to far-infrared plasmon offers appealing prospects for tunable stanene-based terahertz nanophotonics \cite{wang2011found,das2011electronic,basov2011elec,kotov2012electron,tame2013quantum,
fei2012gate,grigorenko2012g,bao2012graphene,luo2013plasmons,
jablan2013plasmons,low2014graphene,basov2014colloquium,
xia2014two,stauber2014plas,politano2014plasmon,garcia2014graphene,
xiao2016graphene,thygesen2017cal,huang2017graphene,fan2019graphene,
rodin2020collective,li2020graphene,in2020dirac,elbanna20232d,kuang2024optical}. The following discussion therefore focuses on the low-energy plasmon near $\Gamma$ point, where the effects of temperature, chemical potential, and a perpendicular electric field on this collective plasmon mode are examined. We note that, the wave vector $\boldsymbol{q}$ of the low-energy plasmon of stanene follows the M--$\Gamma$ (left) and $\Gamma$--K (right) path, and this convention is adopted for all the following figures and is not repeated in separate figure captions.

\subsection{Temperature dependence of low-energy plasmon}

\begin{figure*}[t]
	\centering
	\includegraphics[width=16cm]{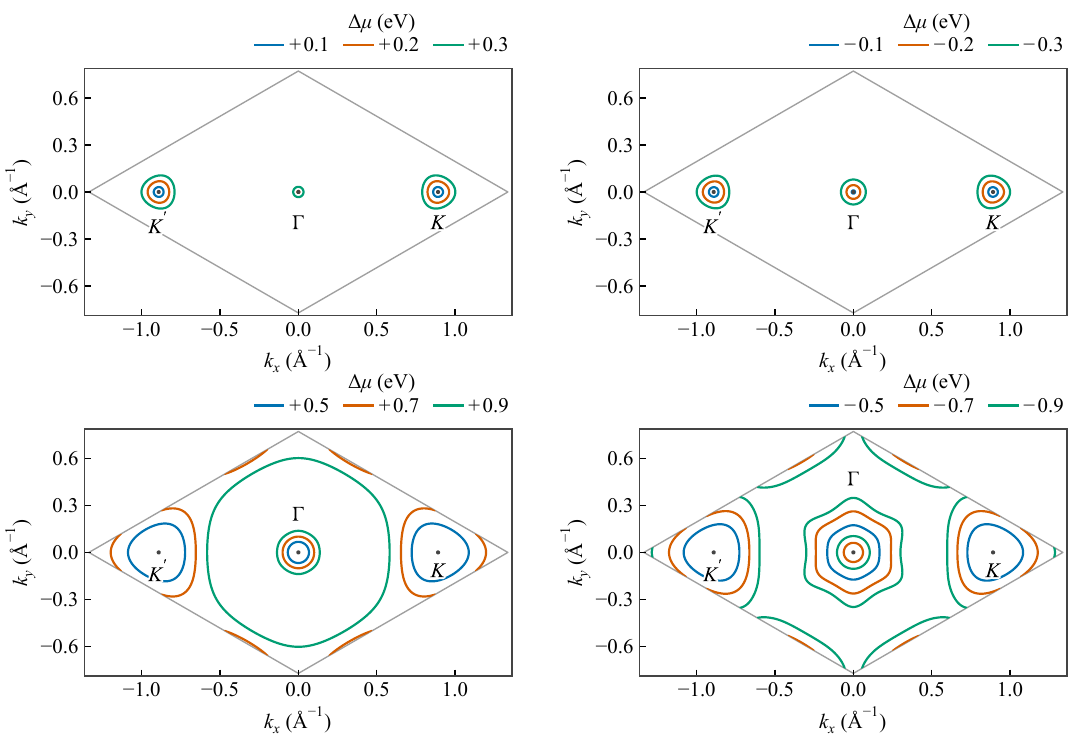}
	\caption{Fermi surface profiles of stanene under various chemical-potential shifts $\Delta\mu$ for electron and hole dopings.
	}
	\label{fig:fm}
\end{figure*}

Figure~\ref{fig:temperature_spectra} presents the low-energy dynamical response function and the energy-loss function of charge-neutral monolayer stanene at $T=250$, $350$, and $500~\mathrm{K}$. At all examined temperatures, a bright, well-defined plasmon ridge emerges near the $\Gamma$ point at low energy, exhibiting the characteristic $\sqrt{q}$ dispersion of 2D low-energy plasmons. With temperature increased sequentially from 250 to 350 and then to $500~\mathrm{K}$, the plasmon exhibits three temperature-induced behaviours: broadening of its energy-wavevector range, an upward energy shift of the plasmon dispersion at fixed momentum $q$, and a pronounced enhancement of the loss-function resonance intensity. These stable low-energy plasmon modes lie outside the Lindhard continuum, where Landau damping is quite weak. With increasing temperature, single-particle excitations are also enhanced, expanding the region over which Landau damping becomes significant.

To characterize the temperature evolution of low-energy plasmon in stanene more quantitatively, we calculate the plasmon peak energy $E_p$, peak height $L_p$, normalized spectral linewidth $\mathrm{FWHM}/(2\eta)$, and quality factor $Q_{\boldsymbol q}$ inside the temperature-momentum parameter space and plot them in Fig.~\ref{fig:temperature_phase}. For each $(q,T)$ point, we search for local maxima of the loss function spectrum inside a predefined low-energy window. A candidate loss peak is accepted if a zero crossing of $\operatorname{Re}\varepsilon(q,\omega)$ lies within the energy interval spanned by its full width at half‑maximum. When multiple candidate peaks satisfy this dielectric criterion within the considered energy window, we select the peak with the maximum amplitude as the target plasmon resonance. For the parameter regime investigated in this work, the low-energy plasmon of interest corresponds to the most prominent loss peak among these candidates. This extraction scheme enables us to consistently track the low-energy collective excitation in the $(q,T)$ parameter space. We also note that the exact $q=0$ point is not used directly in the numerical calculations, since the Fourier-transformed Coulomb interaction $V(\boldsymbol q)=e^2/(2\varepsilon_0\varepsilon_r q)$ diverges as $1/q$ in this limit. Instead, the $\Gamma$ point is represented by a small finite wave vector of magnitude $\sim 1\times10^{-4}$ {\AA}$^{-1}$, which effectively approximates the long-wavelength limit.

The upward shift of plasmon peak energy $E_p$ with increasing temperature in Fig.~\ref{fig:temperature_spectra} originates from thermal population of carriers near the Fermi level. For charge-neutral monolayer stanene, finite temperature excites electrons across the band gap and generates thermally activated electron–hole pairs. This increases the effective free-carrier density, which directly raises the plasmon resonance energy. The temperature–momentum map of $E_p$ in Fig.~\ref{fig:temperature_phase}(a) demonstrates that the energy shift is most pronounced at larger $q$. In the small $q$ regime near the $\Gamma$ point, the peak energy remains comparatively insensitive to thermal excitation, owing to the long-wavelength nature of the collective charge oscillation.

\begin{figure*}[t]
	\centering
	\includegraphics[width=\textwidth]{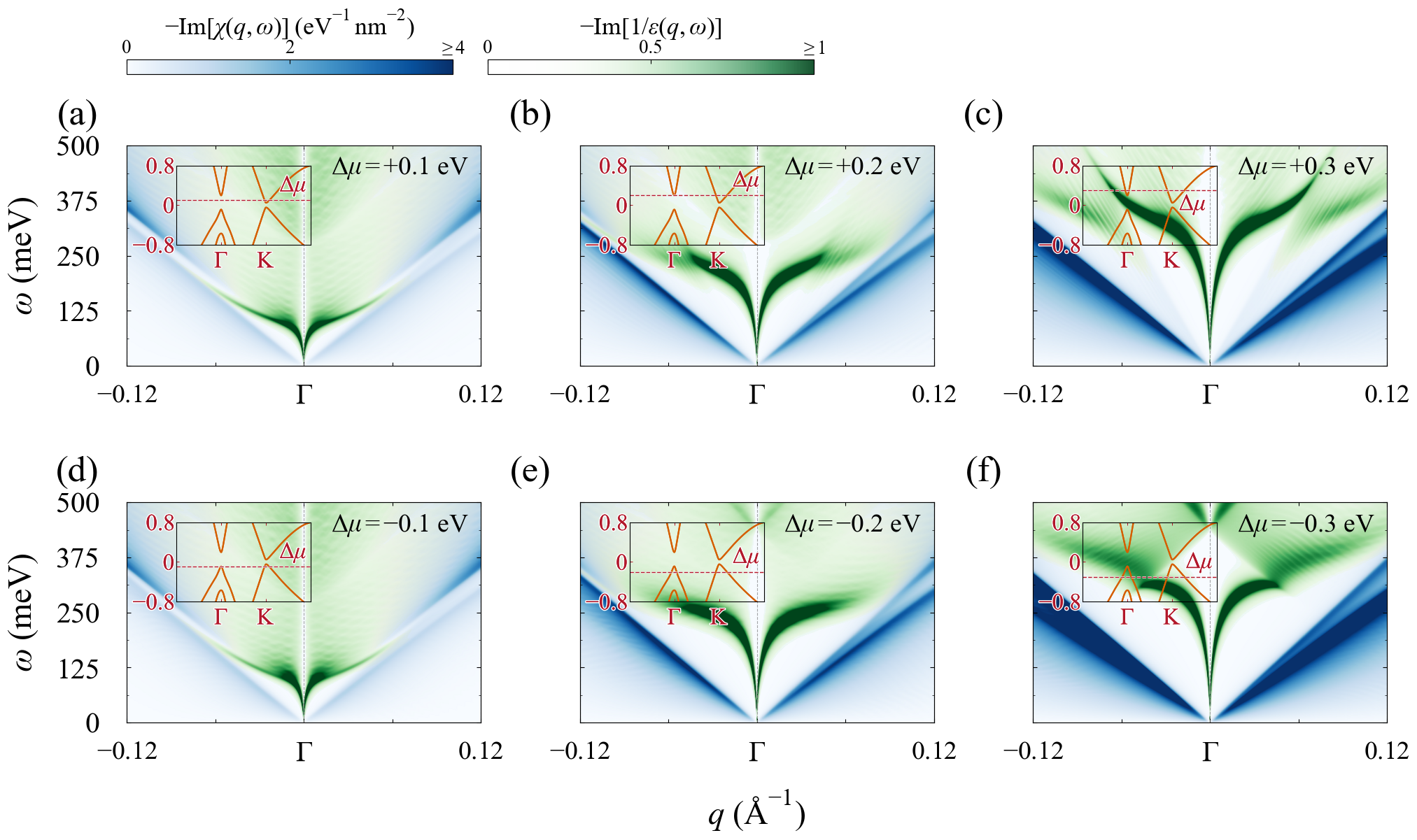}
	\caption{
    Dynamical response and energy-loss function of stanene in moderate-dope cases at $T=300~\mathrm{K}$ for $\Delta\mu = +0.1$ eV in (a), $\Delta\mu = +0.2$ eV in (b), $\Delta\mu = +0.3$ eV in (c), $\Delta\mu = -0.1$ eV in (d), $\Delta\mu = -0.2$ eV in (e), and $\Delta\mu = -0.3$ eV in (f), with their corresponding insets denoting the low-energy bands involving both $\Gamma$ and $\mathrm{K}$ valleys.
	}
	\label{fig:mu_ctr}
\end{figure*}

Figure~\ref{fig:temperature_phase}(b) presents the temperature-momentum map for the peak height $L_p$ extracted from the energy-loss function. Notably, the enhancement of resonance-peak amplitude upon heating is pronounced in the small $q$ region near the $\Gamma$ point, while this thermal enhancement becomes relatively weaker at larger wave vectors. Within the Lindhard formalism, the density polarization $\chi(\boldsymbol q,\omega)$ is governed by Fermi occupation-factor differences $f(E_{n\boldsymbol{k}})-f(E_{m,\boldsymbol{k+q}})$. As temperature increases, thermal broadening of the Fermi-Dirac distribution modifies these occupation differences and activates additional intraband transitions contributing to the collective density response in the long-wavelength small $q$ regime \cite{DasSarma2013,Patel2015,Gumbs2017}. As shown in Fig.~\ref{fig:temperature_phase}(c), in the long-wavelength small $q$ regime, the normalized spectral linewidth $\mathrm{FWHM}/(2\eta)$ is about $1$, indicating the sharp resonance peak and negligible‌ Landau damping. At larger $q$, a slight deviation of this ratio from $1$ indicates a broadening of the plasmon peak, consistent with the slightly enhanced Landau damping.

Fig.~\ref{fig:temperature_phase}(d) displays the $\log_{10}(Q_{\boldsymbol q})$ map of the plasmon quality factor $Q_{\boldsymbol q}$, which characterizes the relative damping and spectral coherence of the collective plasmon mode. A larger value of $Q_{\boldsymbol q}$ corresponds to weaker relative damping and a larger number of oscillation cycles before significant decay. Overall, throughout the considered temperature range, most of the quality factor $Q_{\boldsymbol q}$ exhibits relatively larger value indicating remarkable collective plasmon modes. Especially in the small $q$ regime, $Q_{\boldsymbol q}$ reaches more than 1000, indicating sharply defined plasmon resonances with suppressed Landau damping and robust collective charge-density polarization response.

In short, with increased temperature, the thermally activated carrier population for plasmon collective oscillations rises, and the energy-wavevector range is expanded for plasmons. Especially, the increased energy shift is pronounced at larger $q$, and the loss-function resonance is remarkably enhanced for the small $q$ regime. Throughout the considered temperature range about from 250 to 600 $\mathrm{K}$, the quality factor has large value more than 100 indicating sharp stable plasmon resonances. The energy of thermally tunable plasmons lies in the terahertz-to-far-infrared regime. Accordingly, temperature-tunable plasmons in charge-neutral monolayer stanene hold promising applications for tunable terahertz nanophotonics.

\subsection{Chemical-potential dependence of the low-energy plasmon}

The doping dependence of plasmon dispersion constitutes the central feature of gate-tunable collective response in monolayer stanene. Upon electrostatic doping via the back gate, the chemical-potential shift $\Delta\mu$ directly controls the Fermi-level position relative to the charge-neutral point, which in turn changes the occupied carrier states inside the Fermi sea for both electron doping $\Delta\mu > 0$ and hole doping $\Delta\mu < 0$. Fig.~\ref{fig:fm} shows the Fermi surface profiles for moderately doped cases with $\Delta\mu = \pm 0.1, \pm 0.2, \pm 0.3$ and heavily doped cases with $\Delta\mu = \pm 0.5, \pm 0.7, \pm 0.9$. For moderately doped situations the Fermi surface is mainly contributed by the Dirac bands near the $\mathrm{K}$ or $\mathrm{K}^\prime$ points, while for heavily doped cases bands from both $\mathrm{K}$ and $\Gamma$ valleys play a comparable role in the Fermi surfaces. Furthermore, with the increased absolute values of $\Delta\mu$, the breaking of particle-hole symmetry becomes more pronounced.

\begin{figure*}[t]
	\centering
	\includegraphics[width=\textwidth]{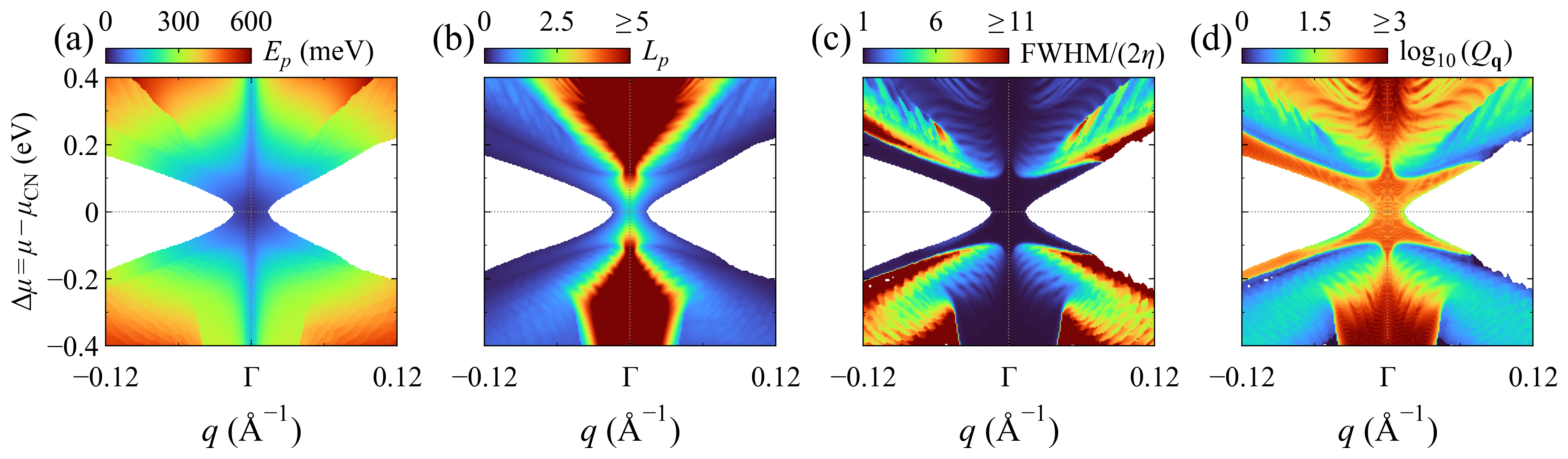}
	\caption{
     $(\mathbf{q}, \Delta\mu)$ maps of the plasmon peak energy $E_p$, peak height $L_p$, normalized linewidth $\mathrm{FWHM}/(2\eta)$, and quality factor $Q_{\boldsymbol q}$ in (a)-(d), respectively, for moderate-doped stanene, at $T=300~\mathrm{K}$.
	}
	\label{fig:mu_phase}
\end{figure*}

\begin{figure*}[t]
	\centering
	\includegraphics[width=\textwidth]{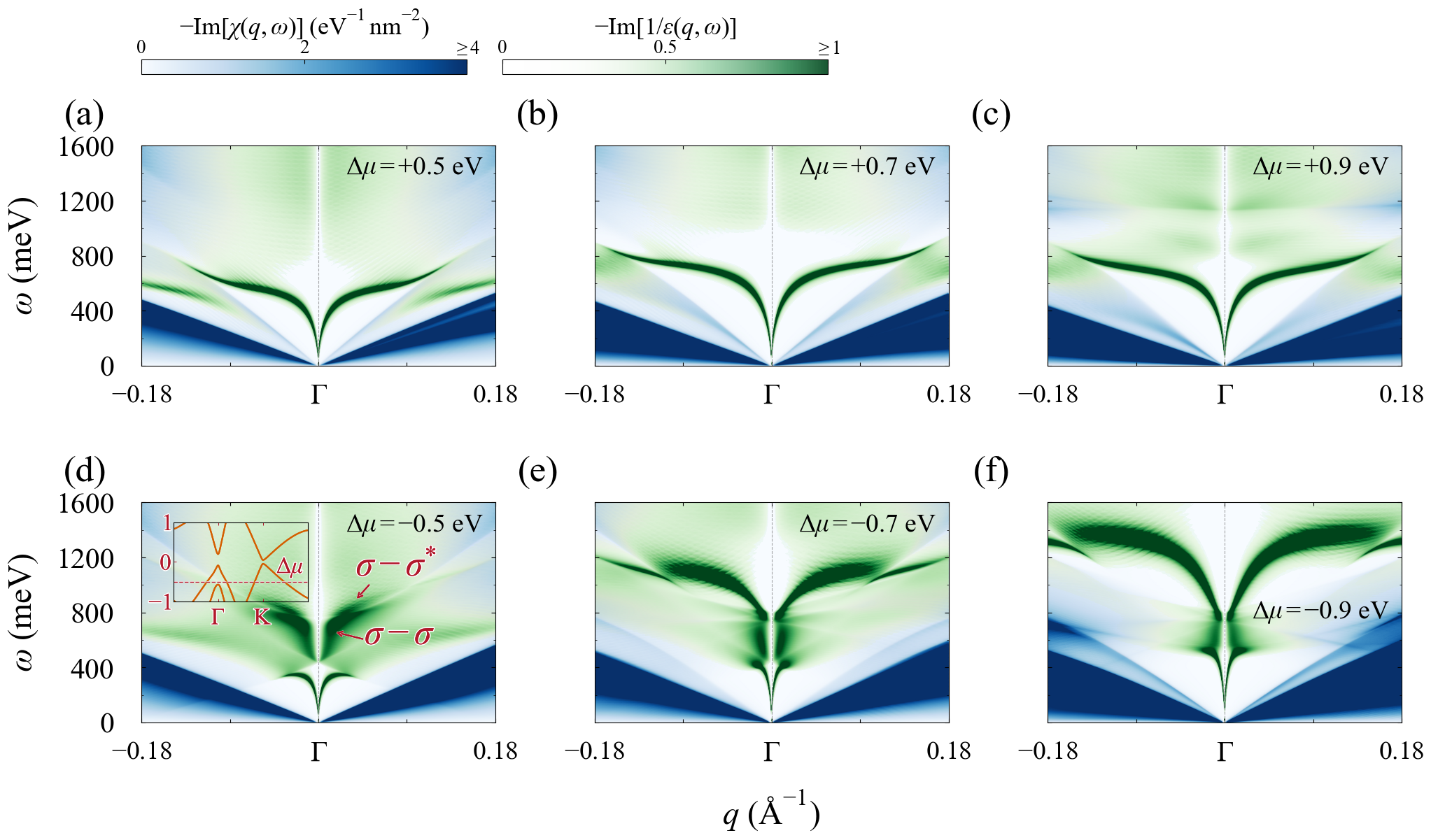}
	\caption{
    Dynamical response and energy-loss function of stanene in heavy-doped stanene at $T=300~\mathrm{K}$ for $\Delta\mu = +0.5$ eV in (a), $\Delta\mu = +0.7$ eV in (b), $\Delta\mu = +0.9$ eV in (c), $\Delta\mu = -0.5$ eV in (d), $\Delta\mu = -0.7$ eV in (e), and $\Delta\mu = -0.9$ eV in (f). The inset in (d) represents the low-energy band structure involving both $\Gamma$ and $\mathrm{K}$ valleys.
	}
	\label{fig:mu_hev}
\end{figure*}

\begin{figure*}[t]
	\centering
	\includegraphics[width=\textwidth]{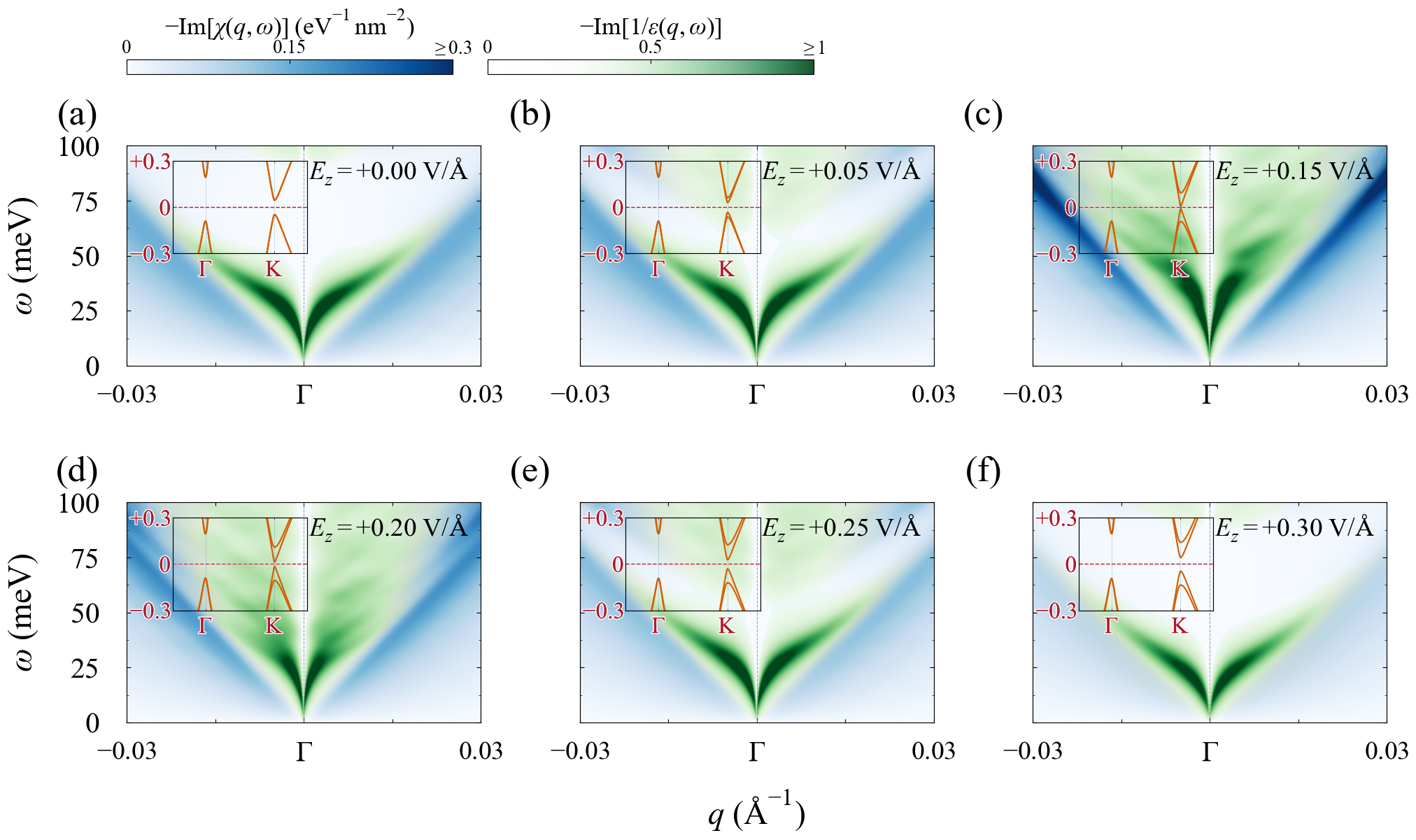}
	\caption{Dynamical response and energy-loss function of charge-neutral stanene at $T=300~\mathrm{K}$ under
            different perpendicular electric fields for $E_z=0$ V/{\AA}  in (a), $E_z= +0.05$ V/{\AA} in (b), $E_z= +0.15$ V/{\AA} in (c), $E_z= +0.2$ V/{\AA} in (d), $E_z= +0.25$ V/{\AA} in (e), and $E_z= +0.3$ V/{\AA} in (f), with corresponding insets denoting their low-energy band structures. 
	}
	\label{fig:field_ctr}
\end{figure*}

Figure ~\ref{fig:mu_ctr} presents the dispersion maps of low-energy dynamical response and energy-loss function under various moderate-doped chemical-potential shifts including $\Delta\mu = \pm 0.1, \pm 0.2$, and $\pm 0.3$ at $T=300 \mathrm{K}$. As the absolute value $|\Delta\mu|$ of the chemical-potential shift increases, the whole plasmon dispersion curve shifts toward higher excitation energy for both electron-doped and hole-doped cases. This means that for a fixed in-plane momentum transfer $\boldsymbol q$, the plasmon resonance energy systematically increases. Physically, raising $|\Delta\mu|$ increases the number of carriers participating in collective charge oscillations, thereby strengthening the collective charge response and shifting the plasmon resonance to higher frequencies. In the small-$q$ regime near the $\Gamma$ point, the plasmon exhibits the characteristic $\omega\propto\sqrt{q}$ and the approximately symmetrical dispersions for $\Delta\mu = \pm 0.1$ in Figs. ~\ref{fig:mu_ctr}(a) and~\ref{fig:mu_ctr}(d), and $\pm 0.2$ in Figs. ~\ref{fig:mu_ctr}(b) and~\ref{fig:mu_ctr}(e), due to the approximate particle-hole band symmetry near the charge-neutral point. For $\Delta\mu = \pm 0.3$, the plasmon dispersions deviate from the $\sqrt{q}$ form in higher energy and larger $q$ regions due to the appearance of $\Gamma$ bands, as denoted by the insets in Figs.~\ref{fig:mu_ctr}(c) and~\ref{fig:mu_ctr}(f). In addition, a new $\sigma$-$\sigma$ plasmon mode near $\Gamma$ point emerges with energy larger than about 0.45 eV. The asymmetry of plasmons between $+0.3$ and $-0.3$ appears because of the broken particle-hole symmetry. Although the single-particle continuum also expands when $|\Delta\mu|$ grows, the collective plasmon mode remains identifiable and persists to larger $q$.

Figure~\ref{fig:mu_phase} presents $E_p$, $L_p$, $\mathrm{FWHM}/(2\eta)$, and $Q_{\mathbf q}$ over a broader parameter range of
$(q,\Delta\mu)$ at $T=300~\mathrm{K}$. From Fig.~\ref{fig:mu_phase}(a) we can see that with increased $|\Delta\mu|$ the plasmon peak energy $E_p$ increases and the range of $\boldsymbol q$ allowing plasmons expands, which is consistent with the results in Fig.~\ref{fig:mu_ctr}. According to Figs.~\ref{fig:mu_phase}(b)-~\ref{fig:mu_phase}(c), the stable plasmon resonances with weak Landau damping mainly lie inside central hourglass-shaped momentum range with larger part for positive $\Delta\mu$. In addition, for relatively large chemical potential shift about $|\Delta\mu|>0.2$, the plasmons between positive and negative $\Delta\mu$ exhibit remarkable asymmetry.

Figure~\ref{fig:mu_hev} shows the dispersion maps of low-energy dynamical response and energy-loss function under various heavy-doped chemical-potential shifts including $\Delta\mu = \pm 0.5, \pm 0.7$, and $\pm 0.9$ at $T=300 \mathrm{K}$. For heavily electron-doped cases with $\Delta\mu = 0.5, 0.7$, and $0.9$ in Figs.~\ref{fig:mu_hev}(a)-~\ref{fig:mu_hev}(c), the total intraband polarizability is the sum of contributions from $\Gamma$ and $\mathrm{K}$ valleys, and their superposition still yields single Drude-type response, supporting the main characteristic $\sqrt{q}$ plasmon branch. With increased $\Delta\mu$, the intraband polarization enhances, and the plasmon dispersion shifts globally towards higher energy. In heavily hole-doped cases, the valence bands with evolved two subbands near $\Gamma$ point are apparently asymmetrical to the single conduction band. Therefore, the heavily hole-doped plasmon will exhibit remarkably different dispersion features. At $\Delta\mu = -0.5$, the topmost valence band is partially populated, supporting conventional intraband transitions, and the second valence band is fully occupied, allowing a transition between the two valence subbands with a threshold of about 0.4 eV. This inter-subband transition enables a new interband $\sigma$-$\sigma$ plasmon mode, and the interband transition between the conduction band and the second valence subband induces another $\sigma$-$\sigma^*$ plasmon mode, as denoted in Fig.~\ref{fig:mu_hev}(d). With heavier doping ($\Delta\mu = -0.7, -0.9$ eV), besides the intraband transitions from the two partially populated valence subbands, the above two interband transitions introduce resonant structures in the real part of the polarizability near their threshold energies, giving rise to multiple solutions of $\mathrm{Re}\varepsilon=0$, and hence complex multi-branch plasmon structures appear in Figs.~\ref{fig:mu_hev}(e) and \ref{fig:mu_hev}(f). The heavier doping also enhances damping, and consequently the plasmon peak at higher energy and larger $q$ is broadened, instead of forming sharp lines.

\begin{figure*}[t]
	\centering
	\includegraphics[width=\textwidth]{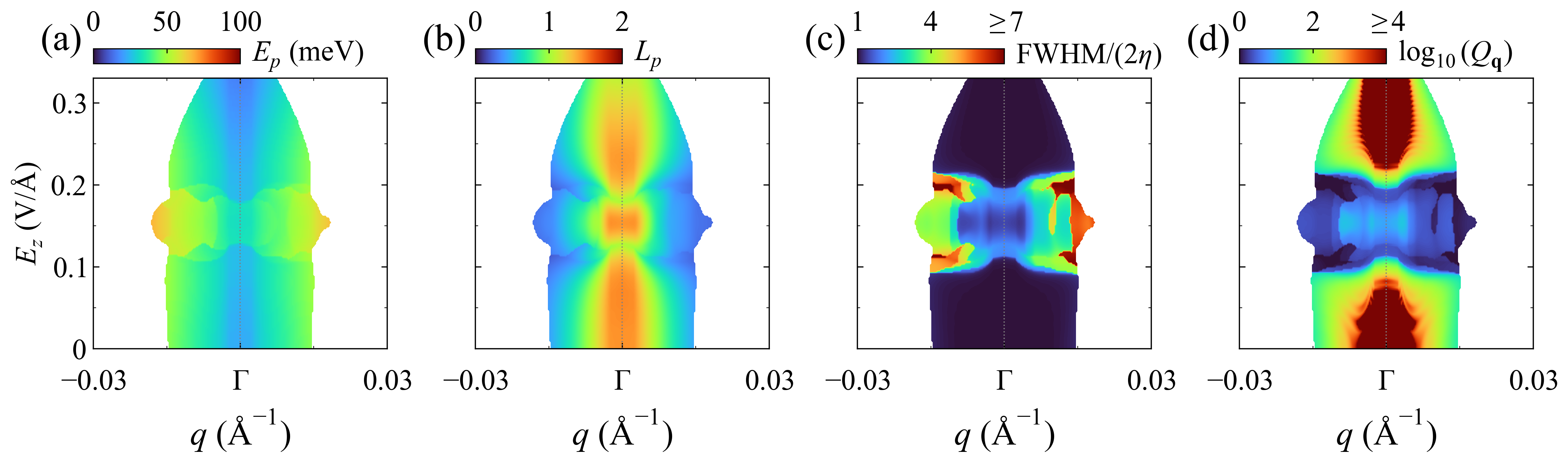}
	\caption{
		$(\mathbf{q}, E_z)$ maps of the plasmon peak energy $E_p$, peak height $L_p$, normalized linewidth
        $\mathrm{FWHM}/(2\eta)$, and quality factor $Q_{\mathbf{q}}$ in (a)-(d), respectively, at $T=300~\mathrm{K}$. For each $E_z$, the corresponding charge-neutral chemical potential is determined separately.
	}
	\label{fig:field_phase}
\end{figure*}

Overall, by applying electrostatic back-gate voltage to induce chemical-potential shift $\Delta\mu$, the carrier occupation inside the Fermi sea can be rearranged and new interband transitions are opened via electrical means. In moderately doped cases, the Dirac bands near the $\mathrm{K}$ point mainly contribute the intraband $\sqrt{q}$ Dirac plasmon. As $|\Delta\mu|$ increases from 0.1 to 0.3 eV, the sheet carrier density grows and the collective charge response is strengthened, which increases the Dirac plasmon resonance energy. In heavily electron-doped cases, the superposition of $\Gamma$ and $\mathrm{K}$ valleys yields a single Drude-type response, maintaining the $\sqrt{q}$ plasmon branch, and its resonance energy can be enlarged by increasing the doping concentration. In heavily hole-doped cases, the interband transitions between two valence subbands and between the valence subbands and the conduction band near $\Gamma$ point induce new $\sigma$-$\sigma$ and $\sigma$-$\sigma^*$ interband plasmon modes or complex multi-branch plasmon structures, giving rise to apparent linewidth broadening of the plasmon resonance. Gate-tunable chemical-potential shift is not only a highly efficient electrical-control approach on plasmon, but also a good probing method on the inner subband structures and collective coherent plasmon structures in stanene.

\subsection{Electric-field dependence of the low-energy plasmon}

We next examine the effect of a perpendicular electric field on the low-energy plasmon in charge-neutral stanene. We first discuss the electric-field effect on the low-energy band structures of stanene. As shown in Fig.~\ref{fig1}(c), the opposite on-site sublattice energy shifts induced by $E_z$ lift the band degeneracy near the $\mathrm{K}$ point;  that is, the original degenerate conduction and valence bands are correspondingly split into two separate branches. As $E_z$ increases, the initial band gap first shrinks to zero and subsequently reopens and rises in Fig.~\ref{fig1}(d). The modified energy bands near the $\mathrm{K}$ point alter the collective plasmon resonance via temperature-driven redistribution of electronic occupations, which enables electric-field tunability of low-energy plasmons in stanene.

Figure~\ref{fig:field_ctr} shows the dispersion maps of low-energy dynamical response and energy-loss function for different
values of $E_z$. As $E_z$ grows from 0 to 0.15 V/{\AA} in Figs.~\ref{fig:field_ctr}(a)-\ref{fig:field_ctr}(c), the initial degenerate bands are split and the band gap shrinks toward zero. Thermal excitation generates a higher density of electron-hole pairs, supplying a substantial carrier population for plasmon collective oscillations. Consequently, the conventional $\sqrt{|q|}$-like plasmon near the $\Gamma$ point extends toward higher energy. Because of the $E_z$-induced band-splitting, several additional plasmon branches deviating from the $\sqrt{|q|}$ dispersion emerge \cite{van2014spin,wu2015effect}, in relatively higher energy regime above approximately 30 meV. When $E_z$ is further increased from 0.2 to 0.30 V/{\AA} in Figs.~\ref{fig:field_ctr}(d)-\ref{fig:field_ctr}(f), the band gap reopens and widens. The density of thermally generated carriers decreases, and the overall plasmon dispersion shifts toward lower energies. One may infer that a further increase of $E_z$ will further enlarge the band gap and deplete thermally excited carriers, so that no low-energy plasmon modes can survive.

Figure~\ref{fig:field_phase} further shows $E_p$, $L_p$, $\mathrm{FWHM}/(2\eta)$, and $Q_{\mathbf{q}}$ over the $(q,E_z)$ parameter space at $T=300~\mathrm{K}$. First, as $q$ moves away from the $\Gamma$ point, several higher-energy plasmon branches show markedly distinct behaviors from the conventional $\sqrt{|q|}$-like plasmon across the four subplots as a result of band splitting. These relatively high-energy plasmon modes exhibit considerably larger linewidths and lower quality factors. Second, outside the range 0.1 V/{\AA} $< E_z <$ 0.2 V/{\AA}, the conventional $\sqrt{|q|}$ plasmon not only possesses favorable characteristics including sharp resonance peaks, narrower linewidths and higher quality factors, but also permits continuous tunability via $E_z$. Third, for $E_z>0.25$ V/{\AA}, increasing $E_z$ widens the band gap, which depletes thermally-generated carriers. Accordingly, the momentum window that sustains plasmon excitations shrinks.

In brief, under temperature-driven electronic occupations, the $E_z$ modulates the low-energy plasmons of charge-neutral monolayer stanene via band-structure reconstruction effects on electronic redistribution rather than extrinsic doping. Our numerical results indicate that, for 0 V/{\AA} $< E_z <$ 0.3 V/{\AA}, $E_z$ permits continuous tunability of the $\sqrt{|q|}$ plasmon. In addition, around the zero band gap, the $E_z$-induced band-splitting enables several new plasmon branches.

\section{Conclusions}
\label{Conclusions}

In summary, using the TB model including spin-orbit coupling and sublattice stagger potentials induced by perpendicular electric field, finite-temperature multiband Lindhard function, and random-phase approximation, we systematically investigate the high-energy plasmon structures as well as the effects of temperature, chemical potential, and perpendicular electric field on the dynamical response and low-energy collective plasmon spectrum of monolayer stanene. 

Four distinct high-energy plasmon modes are resolved in stanene across about 2-4 eV energy window, namely the $\pi$, $\sigma$-$\pi^*$, $\pi$-$\sigma^*$ and $\sigma$-$\sigma^*$ plasmons with well-defined energy positions. Unlike graphene and silicene that support the conventional $\pi$ and $\pi+\sigma$ plasmon modes, stanene possesses two extra high-energy plasmon branches, highlighting its unique high-energy plasmonic properties beyond other group-IV 2D counterparts.

Rising temperature boosts the density of thermally populated carriers contributing to collective plasmon oscillations and extends the energy-wavevector region available to plasmon excitations. A prominent energy shift occurs at large wavevector $q$, while the loss-function resonance undergoes considerable enhancement for small $q$. Over the investigated temperature interval spanning roughly 250–600 $\mathrm{K}$, the low-energy plasmon exhibits high quality factor and robust collective coherence. 

Electrostatic back-gate voltages introduce a chemical-potential shift $\Delta\mu$, which enables carrier redistribution inside the Fermi sea and supports several interband transitions through all-electrical tuning. In moderately doped cases, the Dirac bands near the $\mathrm{K}$ point are mainly responsible for the intraband $\sqrt{q}$ Dirac plasmon. In heavily electron-doped cases, both $\Gamma$ and $\mathrm{K}$ bands concurrently produce the Drude-type intraband response, maintaining the $\sqrt{q}$ plasmon branch, and its resonance energy can be obviously enlarged by increasing the doping concentration. In heavily hole-doped cases, several interband transitions occur near the $\Gamma$ point, including transitions between valence subbands and those from valence subbands to the conduction band, yielding new $\sigma$-$\sigma$ and $\sigma$-$\sigma^*$ interband plasmon modes or complex multi-branch plasmon structures.

For thermally populated electronic states, the perpendicular electric field $E_z$ tunes low-energy plasmons in charge-neutral monolayer stanene by reconstructing the electronic band structure and reorganizing carrier distributions. Our numerical calculations reveal that $E_z$ yields continuous adjustability for the $\sqrt{q}$ plasmon provided 0 V/{\AA} $< E_z <$ 0.3 V/{\AA}. Additionally, in the vicinity of the vanishing band-gap point, $E_z$-induced band splitting generates several additional plasmon branches.

Our results highlight the rich many-body electronic dynamical responses, four high-energy plasmon structures, and low-energy collective excitations in stanene under variable temperature, chemical potential and perpendicular electric field. In addition, these tunable low-energy plasmons in stanene fall into the terahertz-to-mid-infrared spectral region. While retaining good compatibility with conventional field-effect-transistor manufacturing techniques, these tunable plasmons offer great potential for terahertz-to-mid-infrared nanophotonics based on stanene. 

\section{Acknowledgments}
This work is supported by the National Natural Science Foundation of China (Grant No. 12247101), the National Key Research and Development Program of China (Grant No. 2022YFA1402704), the Fundamental Research Funds for the Central Universities (Grant No. lzujbky-2025-jdzx07), the Natural Science Foundation of Gansu Province (No. 22JR5RA389, No.25JRRA799), and the 111 Center under Grant No. B20063.

\bibliography{reference}

\end{document}